\documentstyle[12pt,epsf,epsfig,wrapfig]{article}
\def\pole{{\cal P}_e}

\def\polg{{\cal P}_\gamma}

\begin{document}
\rightline{SLAC-PUB-7320}
\rightline{October 1996}
\rightline{(A/E/I)}
\begin{center}
{\large \bf
The polarized electron beam for the SLAC Linear Collider
\\ }
\vspace{4mm}
{\bf M. Woods}
\\
\vspace{4mm}
{\small
Stanford Linear Accelerator Center\\
Stanford University, Stanford, CA 94309
\\}
\end{center}

\begin{center}

\vspace{4mm}
\begin{minipage}{130 mm}
\small
{\bf Abstract.}
The SLAC Linear Collider has been colliding a polarized electron beam
with an unpolarized positron beam at the $Z^0$ resonance for the SLD experiment
since 1992.  An electron beam polarization of close to 80$\%$ has been achieved
for the experiment at luminosities up to $8 \cdot 10^{29} cm^{-2}s^{-1}$.
This is the world's first and only linear collider, and is a successful 
prototype for the next generation of high energy electron linear colliders.
This paper discusses polarized beam operation for the SLC, and includes
aspects of the polarized source, spin transport and polarimetry.
\end{minipage}
\end{center}
\centerline{{\it Presented at the 12th International Symposium on High
Energy Spin Physics}}
\centerline{{\it Amsterdam, The Netherlands}}
\centerline{{\it September 10-14, 1996}}

\vskip 3.75in
\small
This work was supported in part by Department of Energy contract
DE-AC03-76SF00515
\pagebreak
 
\normalsize
A highly polarized electron beam is a key feature for the current physics
program at SLAC [1].  Fixed target
experiments in End Station A (ESA) study the collision of polarized electrons
with polarized nuclear targets to elucidate the spin structure of the 
nucleon and to provide an important test of QCD.  Using the SLAC Linear
Collider (SLC), collisions of polarized electrons with unpolarized
positrons allow precise measurements of parity violation in the 
{\it Z}-fermion couplings and provide a very precise measurement of the 
weak mixing angle.

Polarized electrons are produced [2] by photoemission from a GaAs 
photocathode.  For SLC operation, two Nd:YAG-pumped Ti:sapphire lasers
produce two 2ns pulses separated by about 60ns.  One of these pulses is used
to make electrons for collisions, and the other one is used to make electrons
for positron production.
 
The laser is polarized with a linear polarizer and two Pockels cells as 
shown in Figure 1.  The axes of the linear polarizer and the PS Pockels
cell are along the {\it x,y} axes, while the axes of the CP Pockels cell
are along {\it u,v} ({\it u,v} axes are rotated by $45^{\circ}$ with respect
to {\it x,y}).  This configuration can generate arbitrary elliptical
polarization, and can compensate for phase shifts in the laser transport
optics.  The laser circular polarization, $\polg$, at the GaAs photocathode
is well approximated by 
$ \polg = sin({V_{CP}-V_{CP}^{T} \over 
V_{\lambda/4}^{CP}} \cdot {\pi \over 2}) cos({V_{PS}-V_{PS}^{T} \over 
V_{\lambda/4}^{PS}} \cdot {\pi \over 2}) $,
where $V_{CP}$ and $V_{PS}$ are the Pockels cell
voltages; 
$V_{\lambda /4}$ is the Pockels cell quarterwave voltage;
$V_{CP}^T$ and $V_{PS}^T$ are phase shifts induced by the laser transport
optics. 
To generate circular polarized light, one nominally operates the CP Pockels
cell at its quarterwave voltage and the PS cell at 0 Volts.  Small corrections
to these voltages are needed to compensate for phase shifts in the transport
optics to the photocathode.
A positive HV pulse on the CP Pockels cell produces one helicity, while
a negative HV pulse produces the opposite helicity.  The sign of this
HV pulse is set by a pseudo-random number generator, which updates
at 120 Hz (the electron beam pulse rate). 
 
\begin{figure}
\epsfig{figure=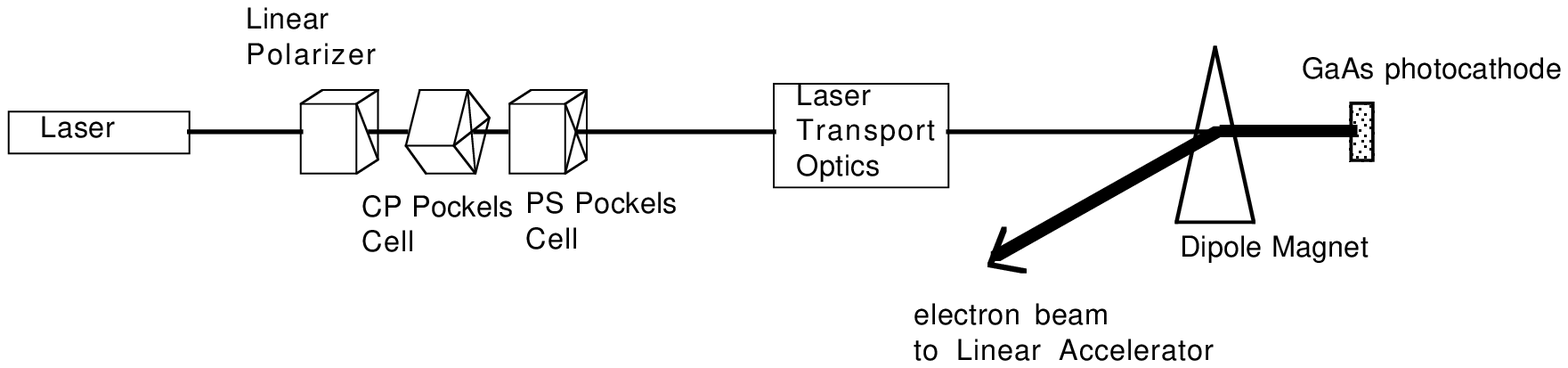,width=15cm}
{\bf Figure 1: SLAC Polarized Electron Source}
\end{figure}

The photoexcitation of electrons in {\it strained} GaAs from
the valence band to the conduction band is illustrated in Figure 2.  The
strain is induced by growing a thin 0.1 $\mu$m layer of GaAs on GaAsP, and
this splits the degeneracy in the j=3/2 valence band by about 50 meV.  
Photons with positive helicity and with energies greater than the band gap
energy of 1.43 eV, but less than 1.48 eV excite the transition shown
from the $m_j=-3/2$ valence level to the $m_j=-1/2$ conduction level.   
The extracted
electrons from the GaAs cathode have the same helicity
as the incident photons since they have opposite
direction to the incident photons (see Figure 1).  In principle, one should
achieve an electron beam polarization, $\pole$, of 100$\%$.  
But in practice, we only achieve $\pole \sim 80\%$.  The
achieved value has been observed to depend on the photocathode quantum 
efficiency (QE) and on the thickness of the strained layer.  How to achieve 
polarizations closer to 100$\%$ is an active research area [3].

\begin{wrapfigure}{r}{8cm}
\epsfig{figure=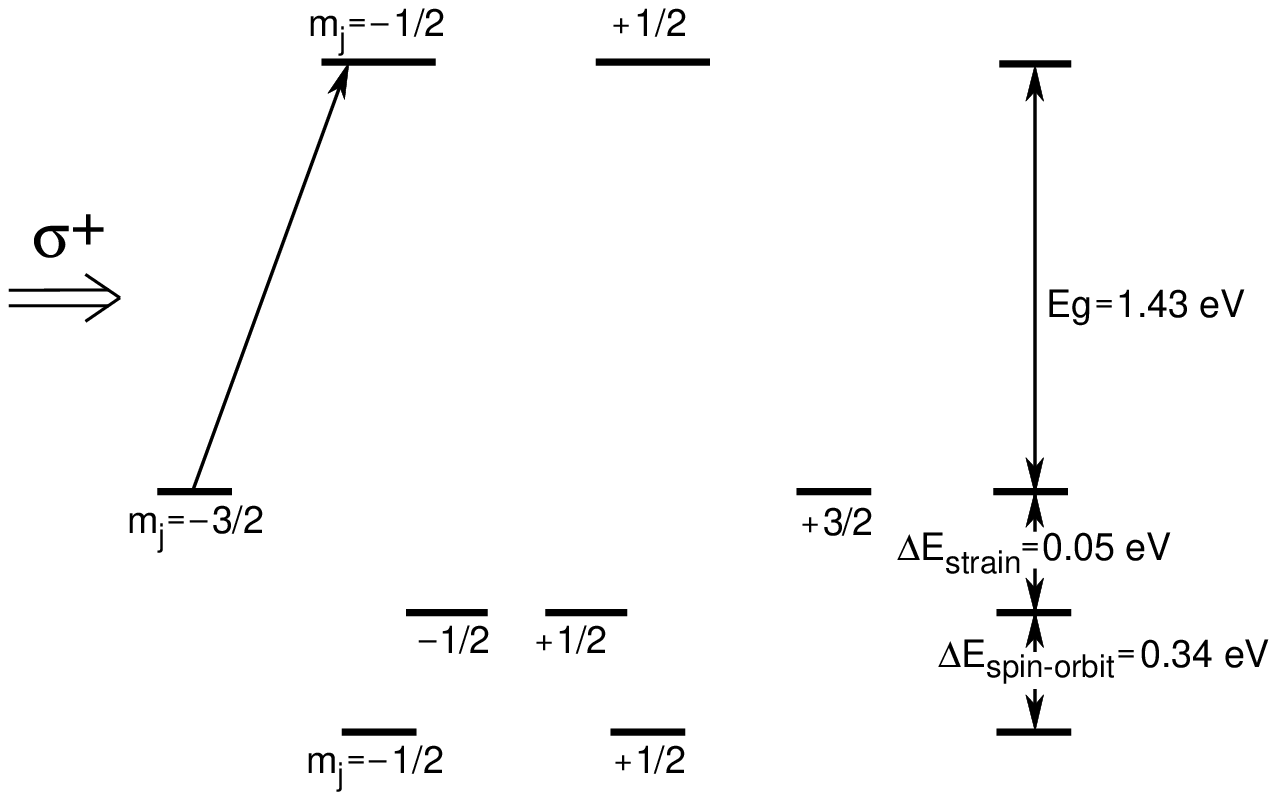,width=8cm}
{\small\bf Figure 2:  GaAs Energy Levels}
\end{wrapfigure}

Two problems encountered with the GaAs photoelectron source are {\it Charge
Limit} [4] and {\it Charge Asymmetry} [5].  Charge Limit is a phenomenon
whereby the instantaneous photocathode QE decreases 
appreciably due to high incident light fluxes, and thereby limits the
maximum electron charge that can be extracted.  Experimentally we find that
the maximum extracted charge is proportional to the QE at low
incident light fluxes and is independent of the maximum laser power available.
In practice, this has not severely impacted the performance of the polarized
electron source at SLAC, but it is an issue that needs to be addressed for
photocathode electron sources.  The Charge Asymmetry problem results from
small linear polarized components in the laser beam and a dependence of the
electron yield on the orientation of the linear polarization.  For laser light
which is 100$\%$ linear polarized, 10$\%$ variations in the electron yield
have been observed depending on the orientation of the linear light 
polarization.  This effect results from anisotropies in the strain [5].  
Even light with $\polg=99.5\%$ has 10$\%$ linear polarization,
and appreciable charge asymmetries between the {\it right} and {\it left}
beam polarization states 
at the $1\%$ level can result.  In practice, this is an easy problem to fix
by monitoring the electron beam charge asymmetry and controlling the Pockels
cell voltages to null the asymmetry.  For the SLAC source, we control the PS
Pockels cell voltage to null the charge asymmetry in a feedback loop. 
It is easy to maintain the
asymmetry below $10^{-4}$ and the level of the average asymmetry is only 
limited by the sophistication of the feedback algorithm and the statistical
fluctuations in the electron charge.
 
Two electron bunches are produced from the photocathode
gun, which operates at 120 kV.  This high voltage is needed to increase the
{\it space charge limit} current capability of the gun above the 6 amps of
peak current needed for SLC operation.  The gun is a complex device where
high voltage arcing and poor vacuum can easily destroy the photocathode QE.
Additionally, the photocathode requires careful preparation to
achieve a negative electron affinity surface to produce an 
adequate QE.  A very large effort was mounted at SLAC to
achieve a robust photocathode electron gun; details of the gun design and
operation can be found in Reference [2].

A schematic of the SLC is shown in Figure 3. 
Two electron bunches from the photocathode gun
are injected into the SLAC Linear Accelerator (Linac)  where they are 
bunched and
accelerated to 1.19 GeV.  They are then kicked by a pulsed magnet
into the Linac-to-Ring (LTR) transfer line to be transported to the
electron
damping ring (DR).  The DR stores the beam for 8ms to reduce the
beam emittance.  The Ring-to-Linac
(RTL) transfer line transports the two bunches from the DR and a pulsed
magnet
kicks them back into the Linac.

\begin{wrapfigure}{r}{5.5cm}
\epsfig{figure=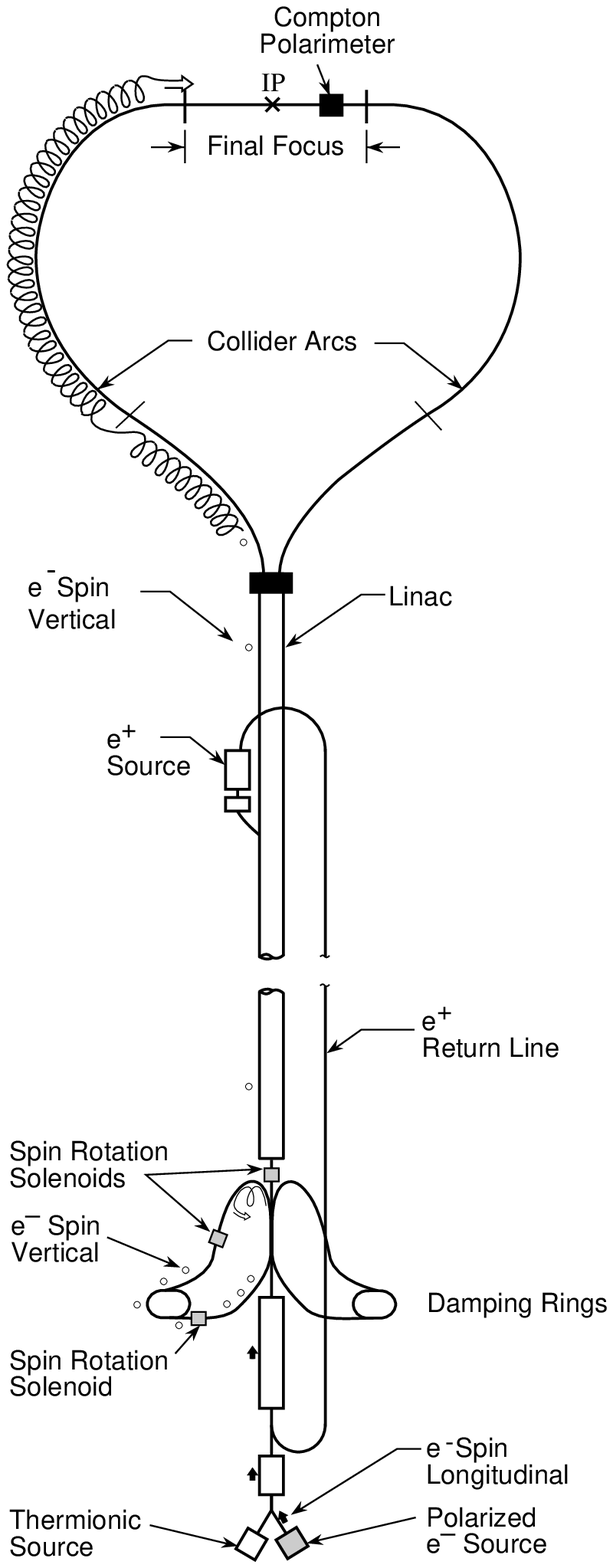,width=5.5cm}
{\small\bf Figure 3:  The Polarized SLC}
\end{wrapfigure}

These two bunches are preceded down the
Linac by a positron bunch which has been extracted from the positron DR.
Three bunches are accelerated down the Linac.  The trailing
electron bunch is accelerated only to 30 GeV, and is sent
to the positron production target.
Positrons in the energy range 2-20 MeV
are collected, accelerated to 200 MeV, and transported to near the start
of the Linac for transport to the positron DR, where they are damped
for 16 ms.  At the end of the Linac, the electron and positron energies
are each 46.6 GeV.
A magnet deflects the electron (positron)
bunch into the north (south) collider arc for transport to the
Interaction Point (IP).  In the arcs, the beams lose about 1 GeV in
energy from synchrotron radiation so that the resulting center-of-mass
collision energy is 91.2 GeV, which is chosen to match the
$Z^0$ mass.  The beam energies are measured with energy
spectrometers with an accuracy of 20 MeV.

The electron spin orientation
is longitudinal at the source and remains longitudinal until the
LTR transfer line to the electron DR.  In the LTR,
the electron spin precesses by
450$^\circ$ to become
transverse at the entrance to the LTR spin
rotator solenoid.  This solenoid rotates the electron spin to
be vertical in the DR to preserve the polarization.
The spin orientation is vertical upon extraction from
the DR; it remains vertical during injection into the Linac and
during acceleration to 46.6 GeV down
the Linac.   
 
The SLC arc
transports the electron beam from the Linac to the IP
and is comprised
of 23 achromats, each of which consists of 20 combined function magnets.
At 46.6 GeV,
the spin precession in each achromat is 1085$^\circ$, while the
betatron phase advance is 1080$^\circ$.  The SLC arc is therefore
operating near a spin tune resonance.  A result of this is that
vertical betatron oscillations in the horizontal bends of the
arc's achromats 
can cause the beam polarization to rotate away
from vertical; this rotation is a cumulative effect in successive
achromats.  (The rotation of the vertical spin component in a given
achromat is simply due to
the fact that rotations in {\it x} and {\it y} do not commute, while
the cumulative effect is due to the spin resonance.)
The resulting spin component in the plane of the arc then
precesses significantly.
 
The arc's spin tune resonance, together with misalignments and
complicated rolls in the arc, result in an inability
to predict the spin transport through the arc.
However, we have two good experimental
techniques for orienting the spin longitudinally at the IP.  First,
using the RTL and Linac spin rotator solenoids one can orient the
electron spin to be along the {\it x}, {\it y}, or {\it z} axis at
the end of the Linac.  The {\it z}-component of the arc's spin transport
matrix can be measured with the Compton polarimeter [6], which
measures the longitudinal electron polarization,
$$P_z^C = R_{zx} \cdot P_x^L + R_{zy} \cdot P_y^L + R_{zz} \cdot P_z^L $$
The experimental procedure is referred to as a 3-state measurement,
and is accomplished by measuring $P_z^C$ for each of {\it x}, {\it y},
or {\it z} spin orientations at the end of the Linac.  
The arc spin rotation matrix elements $R_{zx}, R_{zy},
R_{zz}$ are then determined.  This is sufficient to determine the
full rotation matrix, which is described by three Euler angles.
The matrix R can be
inverted to determine the required spin orientation at the end of
the Linac for the desired longitudinal orientation at the IP.  This 
is achieved with appropriate settings of the RTL
and LINAC spin rotators.

A second method to orient the spin longitudinally at the IP takes
advantage of the arc's spin tune resonance.
A pair of vertical  betatron oscillations ('spin bumps'), each spanning
7 achromats in the last third of the arc, are introduced to rotate the
spin.  The amplitudes of these
spin bumps are empirically adjusted to achieve longitudinal
polarization at the IP.  
Thus, the two spin bumps can effectively replace
the two spin rotators.  For the 1992 SLD run, the spin rotator magnets were
used to orient the electron beam polarization at the IP.  But since 1993,
spin bumps have been used instead because the SLC is now colliding flat
beams to achieve higher luminosity and the spin rotators introduce 
unacceptable {\it x-y} coupling [7].
 
The Compton polarimeter measures the average electron beam polarization,
$\pole$, which
can differ from by a small amount from the luminosity-weighted beam 
polarization, $\pole(1+\xi)$.  The dominant effect is a chromatic one:
the electron beam has a finite energy spread $n(E)$; the spin  
orientation of the electron beam has an energy dependence $\theta_S(E)$;
and chromatic aberrations in the final focus result in a 
luminosity dependence on beam energy, $L(E)$.
The beam energy spread is
monitored by automated wire scans at a high dispersion point.
These are done frequently since the energy distribution of the beam
can easily change.  The dependence of the spin orientation and luminosity on 
the beam energy are more stable.  During the 1994/5 SLD run, 4 
measurements
of $\theta_S(E)$ were made.  Estimates of $L(E)$ are
made by 3 techniques:  simulations of the final focus optics, measurements
of beam spotsizes versus energy, and measurements of the $Z$ boson production
rate for off-energy pulses.  These three estimators give consistent 
results, with the measured data having somewhat less energy dependence
than the simulations.  We use our determinations of $n(E)$, $\theta_S(E)$
and $L(E)$ to estimate the chromatic contribution to $\xi$.  
For the 1994/95 SLD run, we find this contribution to be
$+0.0020 \pm 0.0014$.  An additional effect of similar magnitude
arises from the small precession of the electron spin in the final 
focusing quadrupoles, which for the 1994/95 run contributed 
$-0.0011 \pm 0.0001$ to $\xi$.  We have also measured depolarization 
due to beam-beam effects and find this to be $0.000 \pm 0.001$, 
consistent with theoretical predictions [8].  Combining these 3 effects, we
find $\xi=+0.0009 \pm 0.0017$ for the 1994/95 SLD run.

\vskip 0.15in
The average luminosity-weighted electron beam polarization for the 1994/95
SLD run was $(77.2 \pm 0.5)\%$.
The SLD experiment is approved to continue running with a polarized
electron beam through 1998.  Additionally, there will be a fixed target
experiment, E155, with polarized beam in the spring of 1997.  Future running
of SLD and ESA experiments in parallel with B factory operation at SLAC are
also being considered.

\vfill
{\small\begin{description}
\item{[1]}
M. Woods, {\it AIP Conference Proceedings} {\bf 343}, 230 (1995).  
(Proceedings of SPIN94)
\item{[2]}
R. Alley et al., {\it Nucl. Inst. Meth.} {\bf A365}, 1 (1995).
\item{[3]}
Experimental and theoretical discussions of maximizing the electron 
polarization for GaAs photocathodes were held at the SPIN96 Pre-Symposium
{\it Workshop
on polarized electron sources and low energy polarimeters}; see proceedings.
\item{[4]}
M. Woods et al., {\it J. Appl. Phys.} {\bf 73}, 8531 (1993); H. Tang et al.,
SLAC-PUB-6515 (1994).
\item{[5]}
R.A. Mair et al., {\it Phys. Lett.} {\bf A212}, 231 (1996).
\item{[6]}
M. Woods, SLAC-PUB-7319 (1996), contributed to these proceedings.
\item{[7]}
T. Limberg, P. Emma, and R. Rossmanith, SLAC-PUB-6210 (1993).
\item{[8]}
K. Yokoya, P. Chen, SLAC-PUB-4692 (1988).
\end{description}}
\end{document}